# Faults and Improvements of an Enhanced Remote User Authentication Scheme Using Smart Cards

**Manoj Kumar**

Department of Applied Sciences and Humanities
Hindustan College of Science & Technology
Farah, Mathura, (U.P.) India- 281122.
E. Mail: yamu_balyan@yahoo.co.in

**Abstract** — *In 2000, Hwang and Li proposed a remote user authentication scheme using smart cards to solve the problems of Lamport scheme. Later, Chan- Chang, Shen- Lin- Hwang and then Chang-Hwang pointed out some attacks on Hwang – Li's scheme. In 2003, Shen, Lin and Hwang also proposed a modified scheme to remove these attacks. In the same year, Leung-Cheng-Fong-Chan showed that modified scheme proposed by Shen-Lin-Hwang is still insecure. In 2004, Awasthi and Lal enhanced Shen-Lin-Hwang's scheme to overcome its security pitfalls. This paper analyses that the user U/smart card does not provide complete information for the execution and proper running of the login phase of the Awasthi- Lal's scheme. Furthermore, this paper also modifies the Awasthi- Lal's scheme for the proper functioning* [1].

## I. INTRODUCTION

To authenticate the legitimacy of the remote users over insure channel, a remote password authentication scheme is used. In such a scheme, the password often regarded as a secret shared between the *authentication server* (**AS**) and serves to authenticate the identity of the individual logging on to the server. Through the knowledge of the password, the remote user can use it to create a valid login message to the authentication server. *AS* checks the validity of the login message to provide the access right. Password authentication schemes with smart cards have a long history in the remote user authentication environment. So far different types of password authentication schemes with smarts cards [1] – [3]-[4] –[5] –[6] –[11] –[13] –[14] – [18] have been proposed.

In 1981, Lamport [12] proposed a remote password authentication scheme using a password table to achieve user authentication. In 2000, Hwang and Li [13] pointed that Lamport' s scheme suffered with the risk of a modified password table and the cost of protecting and maintaining the password table. Further, they also proposed a new remote user authentication scheme using smart cards, which is based on the most popular ElGamal's Cryptosystem [15]. This scheme [13] does not maintain the password table to check the validity of the login request. Also, it can withstand message-replaying attack.

In [7], Chan and Cheng pointed out an attack on the Hwang-Li's scheme. In 2003, Shen-Lin-Hwang [9] discussed a different attack on the Hwang-Li's scheme and they also proposed a modified scheme to prevent the attacks on Hwang-Li's scheme. In the same year, Chang and Hwang [3] explained the practical problems of the Chan – Cheng's attack on the Hwang-Li's scheme and Leung – Cheng - Fong and Chen [10] pointed out that the Shen-Lin-Hwang's scheme is still vulnerable to the attack proposed by Chan and Cheng. In 2004, Awasthi - Lal [1] proposed an enhanced remote user authentication scheme using smart cards.

### A. Contributions

This paper aims to show that Awasthi - Lal's scheme required more information for the execution of the login phase**.** This paper also presents a modified remote user authentication scheme to remove the problems of Awasthi - Lal's scheme.

### B. Organization

Section II reviews the Hwang – Li's scheme [13]. Section III describes the cryptanalysis of Hwang – Li's scheme. Section IV reviews the Shen-Lin-Hwang's scheme [9]. Section V reviews the Leung – Cheng - Fong and Chen [10] attack on Shen-Lin-Hwang's scheme. Section VI reviews the Awasthi - Lal's scheme [1]. Section VII yields an objection on the Awasthi - Lal's scheme and pointed out the problem of the scheme. An improved variant of remote user authentication scheme is presented in section VIII. The security of the proposed scheme is discussed in section IX. Finally, comes to a conclusion in the section X.

## II. REVIEW OF THE HWANG-LI'S SCHEME

There are three phases in the Hwang-Li's scheme**:** the registration phase, login phase and the authentication phase. In the registration phase, the user *U* sends a request to the *AS* for the registration. The *AS* will issue a smart card and a password to every user legal through a secure channel. In the login Phase, when the user *U* wants to access the *AS,* she/he inserts her/his smart card to the smart card reader and then keys the identity and the password to access services. In the authentication phase, the *AS* checks the validity of the login request.

### A. Registration Phase

User *U* submits her/his *ID* to the *AS*. *AS* computes the password *PW* for the user *U*, as,

$$PW = ID^{x_s} \mod p,$$

where, $x_s$ is a secret key maintained by the *AS* and *p* is a large prime number. *AS* provides a password *PW* and a smart card to the user *U* through a secure channel. The smart card contains the public parameters (*f*, *p*), where *f* is a one-way function.

### B. Login Phase

User *U* attaches her/his smart card to the smart card reader and keys *ID* and *PW*. The smart card will perform the following operations:
1. Generate a random number *r*.
2. Compute $C_1 = ID^r \mod p$.
3. Compute $t = f(T \oplus PW) \mod p - 1$, where *T* is the current date and time of the smart card reader.
4. Compute $M = ID^t \mod p$.
5. Compute $C_2 = M(PW)^r \mod p$.
6. Sends a login request $C = (ID, C_1, C_2, T)$ to the *AS*.

### B. C. Authentication Phase

Assume *AS* receives the message *C* at time $T_c$, where $T_c$ is the current date and time at *AS*. Then the *AS* takes the following actions:
1. Check the format of *ID*. If the identity format is not correct, then *AS* will rejects this login request.
2. Check, whether $T_c - T \leq \Delta T$, where $\Delta T$ is the legal time interval due to transmission delay, if not, then rejects the login request C.
3. Check, if $C_2 (C_1^{x_s})^{-1} \stackrel{?}{=} (ID)^{f(T \oplus PW)} \mod p$, then the *AS* accepts the login request. Otherwise, the login request will be rejected.

## III. CRYPTANALYSIS OF THE HWANG-LI SCHEME

### A. Chan and Cheng's Attack

According to Chan and Cheng [7], a legal user Alice can easily generate a valid pair of identity and password without the knowledge of secrete key '$x_s$' of *AS*. Alice uses her valid pair ($ID_A$, $PW_A$) to generate another valid pair ($ID_B$, $PW_B$) as follows:

Alice computes $ID_B = (ID_A \times ID_A) \mod p$. Then, she can computes the corresponding password

$$PW_B = ID_B^{x_s} \mod p$$
$$= (ID_A \times ID_A)^{x_s} \mod p$$
$$= (PW_A \times PW_A) \mod p$$

As a result, Alice can generate a valid pair ($ID_B$, $PW_B$) without knowing the secret key $x_s$.

### B. Shen-Lin-Hwang's Attack: Masquerading Attack

According to Shen, Lin and Hwang [9] masquerading attack is possible on Hwang-Li's scheme. A user Bob can masquerade another user Alice to login a remote server and gain access right.

Bob computes an identity $ID_B = ID_A^k \mod p$, where *k* is a random number such that $gcd(k, p) = 1$. Then, he submits this identity $ID_B$ to *AS* for registration. *AS* provides a smart card and a password

$$PW_B = ID_B^{x_s} \mod p.$$

With the knowledge of $PW_B$, Bob can compute

$$PW_A = ID_A^{x_s} \mod p = PW_B^{-k} \mod p.$$

As a result, Bob can masquerade as Alice to login a remote server and gain access privilege.

### C. Chang-Hwang's Attack

According to Chang and Hwang [2], there is a mistake in the Chan-Cheng's attack. It is not always possible that the square of a legal identity satisfies the specific identity format. Chang and Hwang generalized the Chan-Cheng's attack. They described two attacks.

#### 1) Attack-I

Alice computes $ID_B = ID_A^k \mod p$, where *k* is a random number. Then, he can computes the corresponding password

$$PW_B = PW_A^k \mod p.$$

As a result, a legal user Alice can impersonate other user Bob with a valid pair of ($ID_B$, $PW_B$) to login the *AS*. If $ID_A$ is a primitive root of $Zp$, then all the valid identities and their corresponding password can be generated easily.

#### 2) Attack-II

A group of eavesdroppers may cooperate to generate a valid pair of identity ($ID_G$, $PW_G$), as follows:

$$ID_G = \prod ID_{Aj} \mod p \text{ and } PW_G = \prod PW_{Aj} \mod p$$

Chang and Hwang pointed out that in Hwang – Li's scheme, it is still difficult to obtain the corresponding password for a known arbitrary valid identity, but once the valid identity is generated, its corresponding password will be obtained easily.

## IV. SHEN, LIN AND HWANG'S SCHEME

Shen-Lin-Hwang [9] proposed a modified remote user authentication scheme to solve the security pitfall of the Hwang-Li's scheme. Shen-Lin-Hwang's scheme uses the concept of hiding the identity to prevent the masquerading attack. They modified the registration phase, now a shadow identity *SID* will be issued to the legal user. This modified in the registration phase is described below.

### A. Modified Registration Phase

A user *U* submits her/his identity string *J* to the *AS* for the registration. The string *J* contains the name, address, unique number etc. This information in the string *J* is unique for every user. Then the *AS* computes a pair (*SID*, *PW*) for the user *U* after the identity *J* is identified. The pair (*SID*, *PW*) is computed as follows:

$SID = Red\ (J)$ and $PW = (SID)^{x_s} \bmod p$

where, $Red\ (.)$ is a shadow identity of device which is only maintained in the remote server and $S_{ID}$ is the shadow identity of the user $U$. Furthermore, the *AS* distributes the smart card and $(SID, PW)$ to the user $U$ in a secure way. The smart card contains the public parameters $(f, p)$.

In this scheme, the message sent to the *AS* now contains $(SID, C_1, C_2, T)$. Because $J_i$ specially formatted, the evil user cannot compute new identity string $J_i$ via $SID_i$.

## V. CRYPTANALYSIS OF THE SHEN, LIN AND HWANG'S SCHEME

### A. Leung- Cheng-Fong –Chan's Attack

According to Leung-Cheng-Fong-Chan [10], Shen-Lin-Hwang's scheme [9] defends the attacks of registration for a new identity $ID_B$ via $ID_A$ for a legal user Alice. They also pointed out that the modified scheme is still vulnerable to the attack described by Chan and Cheng. They showed that the modified scheme is not secure against the attack that is similar to Chan - Cheng and Chang – Hwang's attacks. If we replace $ID_A$ with $SID_A$, then the Chang and Hwang's attack will work as follows.

Alice computes $SID_B = (SID_A)^k \bmod p$, where $k$ is a random number. Then, he can computes the corresponding password

$$PW_B = PW_A^{\ k} \bmod p.$$

As a result, a legal user Alice can impersonate other user Bob with a valid pair of $(SID_B, PW_B)$ to login the AS. If $SID_A$ is a primitive root of $Z_p$, then all the valid identities and their corresponding password can be generated easily. Since, Chan – Cheng's attack is one case of this attack so it also works well.

## VI. REVIEW OF THE AWASTHI - LAL 'S SCHEME

Awasthi and Lal [1] suggested a new way to enhance the security of the Shen-Lin-Hwang's scheme [9] against Leung-Cheng-Fong-Chan's attack [10]. This section provides a review of Awasthi and Lal's scheme.

### A. Initialization Phase

In this phase, the *AS* generates the following parameters

- $p$ : a large prime number.
- $f$ : a one-way function.
- $x_s$ : a secret key of the system, which is only possessed with the *AS*.
- $Red\ (.)$ : a function which is re-direct the identity of the user and only possessed with the *AS*.

### B. Registration Phase

User $U$ submits her/his identity $ID$ to the *AS*. Then, *AS* computes the followings:

$SID = Red\ (ID)$, and $PW = (SID)^{x_s} \bmod p$.

The *AS* distributes the smart card and the pair $(SID, PW)$ to the user $U$. The smart card contains the public parameters $(f, p)$.

### C. Login Phase

To login at any time $T$, the user $U$ attaches her/his smart card and keys her/his identity $ID$ and password $PW$. The smart card does the following:

1. Generate a random number $r$.
2. Computes $C_1 = (SID)^r \bmod p$.
3. Compute $t = f(T \oplus PW) \bmod p - 1$.
4. Compute $m = (SID)^t \bmod p$.
5. Compute $C_2 = m\ (PW)^r \bmod p$.
6. Sends a message $L_R = (ID, C_1, C_2, T)$ to the *AS*.

### D. Authentication Phase

Assume that the *AS* receives the login request $L_R$ at time $T_S$. Then, *AS* does the following steps:

1. Test the validity of the *ID*.
2. Computes $SID = Red\ (ID)$.
3. Test the validity of the time delay between $T_S$ and $T$.
4. Check, if $C_2 = (C_1^{x_s})(SID)^{f(T \oplus PW)} \bmod p$, then the *AS* accepts the login request. Otherwise, the login request will be rejected by *AS*.

## VII. OUR OBJECTION

This section analyzes the login phase of Awasthi and Lal's scheme. Although Awasthi and Lal suggested a different way to enhance the security of Shen-Lin-Hwang's scheme [9] against Leung-Cheng-Fong-Chan 's attack. But, when we observe the login phase of Awasthi and Lal's scheme, it is found that the user U/smart card does not provide complete information for the execution and proper running of the login phase.

In Awasthi and Lal's scheme to login at any time $T$, the user $U$ attaches her/his smart card to the smart card reader and keys her/his identity *ID* and password *PW*. *In the second steps, the smart card computes $C_1 = (SID)^r \bmod p$, while the number SID neither stored in the smart card nor this number is supplied by the user U at the time of login.*

In the login phase, we observe that the information which are provided by the User $U$ and through the smart for the login to *AS* are not sufficient for the functioning of the login phase. However, the scheme is not fit for the practically implementation.

Consequently, the login phase of Awasthi and Lal's scheme does not work properly due to the lack of information. As a result, Awasthi and Lal's scheme is not suitable for real grounds.

## VIII. AN IMPROVED SCHEME

This section modifies Awasthi and Lal's scheme to attain better and proper functioning. The proposed scheme is an improved variant of the original scheme: *Hwang and Li's scheme*. The modified scheme is described below.

### A. Initialization Phase

In this phase, the *AS* generates the following parameters

- $p$ : a large prime number.

| | | |
|---|---|---|
| $f$ | : | a one-way function. |
| $x_s$ | : | a secret key of the system, which is only possessed with the *AS*. |
| $Red$ (.) | : | a function which is re-direct the identity of the user and only possessed with the *AS*. |

### B. Registration Phase

Assume that this phase is executed over a secure channel. User *U* submits her/his identity *ID* to the *AS*. Then, *AS* computes the followings:

$$S_{ID} = Red\ (ID)\ \text{and}\ PW = (S_{ID})^{x_s} \bmod p.$$

Here $S_{ID}$ is the redirected identity corresponding to the registered identity *ID*. The *AS* distributes the smart card and $(S_{ID} \| ID, PW)$ to the user *U* in a secure way (say physically). The smart card contains the public parameters $(f, p)$.

### C. Login Phase

The user *U* attaches her/his smart card to the smart card reader at any time *T* and keys her/his identity $S_{ID} \| ID$ and password *PW*. The smart card conducts the following computations:

1. Generate a random number *r*.
2. Computes $C_1 = (S_{ID})^r \bmod p$.
3. Compute $t = f(T \oplus PW) \bmod p - 1$.
4. Compute $m = (S_{ID})^t \bmod p$.
5. Compute $C_2 = m\ (PW)^r \bmod p$.
6. Sends a message $L_R = (ID, C_1, C_2, T)$ to the *AS*.

### D. Authentication Phase

Assume that the *AS* receives the login request $L_R$ at time $T_c$. Then, *AS* does the following computations to check the validity of the login request $L_R$.

7. Check the specific format of *ID*. If the format of the *ID* is incorrect, then *AS* rejects the login request $L_R$.
8. Check, whether $T_c - T \leq \Delta T$, where $\Delta T$ is the legal time interval due to transmission delay, if not, then *AS* rejects the login request $L_R$.
9. Computes $S_{ID} = Red\ (ID)$.
10. Check, if $C_2 \stackrel{?}{=} (C_1^{x_s})(S_{ID})^{f(T \oplus PW)} \bmod p$, then the *AS* accepts the login request. Otherwise, the login request will be rejected by *AS*.

## IX. SECURITY ANALYSIS OF THE IMPROVED SCHEME

The above scheme is a modified form of the original scheme: Hwang-Li's scheme. The security analysis has been already discussed and demonstrated in [13]. Therefore, this section will only discuss the enhanced security features of the proposed scheme.

### A. Shen- Lin- Hwang's attack

The functionality of the redirected function *Red* (.) blocks the masquerade attack via registration phase: *Shen- Lin-Hwang's attack*. The function *Red* (·) generates a valid number $S_{ID}$ with the help of the identity *ID*, which is sent by the user *U* at the time of registration request. The eavesdropper cannot compute a valid $S_{ID}$ for any fake identity $ID_F$.

Thus, Bob cannot masquerades as Alice. For the reason, this scheme is secure against the masquerading attack.

### B. Chan- Cheng's attack and Chang- Hwang's Attack

The functionality of the redirected function *Red* (.) also blocks the attacks via authentication phase: *Chan- Cheng's attack, Chang- Hwang's Attack and all extended attacks*. However, Alice computes $S_{ID_B} = (S_{ID_A})^k \bmod p$, where *k* is a random number. Then, he can computes the corresponding password

$$PW_B = PW_A^{\ k} \bmod p.$$

This result is still incomplete; for the success of these attacks, it is essential to obtain the corresponding $ID_B$. In the proposed scheme, the redirected function *Red* (.) is a secret function and always in possession of *AS* only. As a result, a legal user Alice cannot compute a valid pair of redirected identity $S_{ID}$ /identity *ID* and password to impersonate other user Bob to login the *AS*. Thus, *Chang- Hwang's Attack* will not work. Since, *Chan – Cheng* and *Leung – Cheng - Fong and Chen*' s attacks are the extended form of *Chang- Hwang's Attack,* so these attacks also do not work on our scheme.

## X. CONCLUSIONS

This paper analyzed that in Awasthi and Lal's scheme, the *user U/smart card does not provide complete information for the execution and proper running of the login phase.* Furthermore, this paper also proposed a modified remote user authentication scheme, which is secure against *Shen- Lin-Hwang's attack Chan- Cheng's attack and Chang- Hwang's Attack* and all extended attacks.

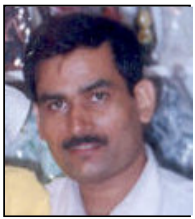 **Manoj Kumar** received the B.Sc. degree in mathematics, in 1993; the M. Sc. in Mathematics, in 1995; the M.Phil., in *Cryptography*, in 1997; the Ph.D. in *Cryptography*, in 2003. He also taught applied Mathematics at DAV College, Muzaffarnagar, India from Sep, 1999 to March, 2001; at S.D. College of Engineering & Technology, Muzaffarnagar, U.P., India from March, 2001 to Nov, 2001; at Hindustan College of Science & Technology, Farah, Mathura, continue from Nov, 2001. He also passed the *National Eligibility Test* (NET), conducted by *Council of Scientific and Industrial Research* (CSIR), New Delhi- India, in 2000. He is a member of Indian Mathematical Society, Indian Society of Mathematics and Mathematical Science, Ramanujan Mathematical society and Cryptography Research Society of India. His current research interests include Cryptography, Numerical analysis, Pure and Applied Mathematics.